# Optimisation of Least Squares Algorithm: A Study of Frame Based Programming Techniques in Horizontal Networks

C. P. E. Agbachi

*Department of Mathematical Sciences, Kogi State University*
*Anyigba, Kogi State, Nigeria*

***Abstract—*** *Least squares estimation, a regression technique based on minimisation of residuals, has been invaluable in bringing the best fit solutions to parameters in science and engineering. However, in dynamic environments such as in Geomatics Engineering, formation of these equations can be very challenging. And these constraints are ported and apparent in most program models, requiring users at ease with the subject matter. This paper reviews the methods of least squares approximation and examines a one-step automated approach, with error analysis, through the instrumentality of frames, object oriented programming.*

## I. INTRODUCTION

In science and engineering studies, regression analysis is a very important tool in providing explanatory models to observed phenomena. There are instances that range from simple linear relationships to rigorous least squares approach.

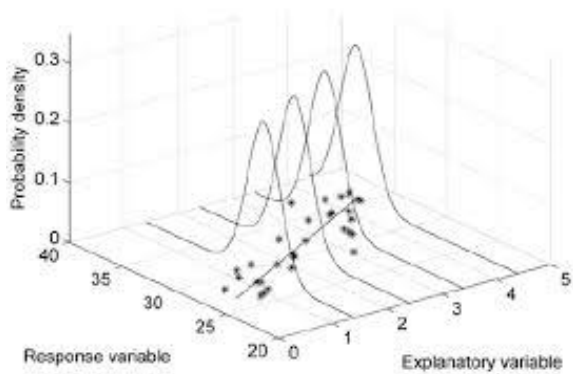

Fig. 1 Scatter Plot

In Fig.1 is a scatter plot from an experiment. The quest given this pattern is to obtain a model that is explanatory and basis for mathematical and scientific projections. This can be done by studying the form of pattern through the concepts of regression analysis.

### A. Simple Regression Analysis

Imagine points on a scatter diagram that assume approximately, the form of a straight line described by $y = bx + a$. Then how a line could be drawn that best fit the data becomes a trial. Intuitively, an acceptable form maybe be obtained, but what is required is mathematical criteria that gives "the best" regression line through the points.

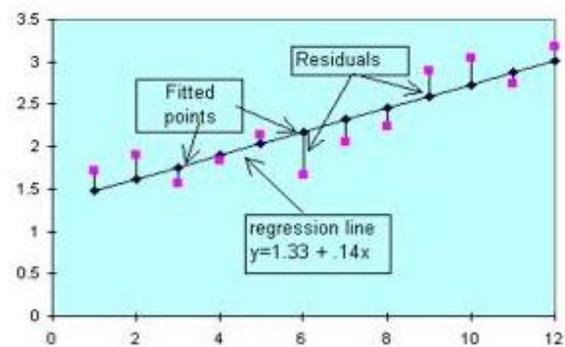

Fig.2 Regression Analysis

A regression line, Fig 2, is simply a single line that fits the data, in terms of having the smallest overall distance from the line to the points. In Statistics this technique for finding the best-fitting line is known as simple linear regression analysis. And by due consideration, the best criterion is that for which the sum of squares of the residuals is a minimum, giving rise to least squares method [1].

### 1. Least Squares Method:

Least squares method of estimation can be used in linear concept to solve the problem of finding a line or curve that best fits a set of data points. Consider the usual case of a set of N pairs of observations $\{X_i, Y_i\}$ where it is required to find a function relating the value of the dependent variable (Y) to the values of an independent variable (X).

In this case, the projection is given by the following equation [2]:

$$\bar{Y} = a + bX$$

In the equation, $\bar{Y}$ is the regression line, while *a* and *b* are the intercept and gradient, respectively. And with reference to Fig. 2, the residual $\vartheta$ has the following expression for N points:

$$\vartheta = \sum_{i=1}^{N}(Y_i - \bar{Y}_i)^2 = \sum_{i=1}^{N}(Y_i - (a + bX_i))^2$$

Given this relationship, the criterion of least squares is enforced through partial differentiation. Thus, in taking the derivative of $\vartheta$ with respect to *a* and *b* and setting them to zero a normal set of equations is obtained:

$$\frac{\partial \vartheta}{\partial a} = 2Na + 2b\sum_{i=1}^{N}X_i - 2\sum_{i=1}^{N}Y_iX_i = 0$$

And

$$\frac{\partial \vartheta}{\partial b} = 2b\sum_{i=1}^{N}X_i^2 + 2a\sum_{i=1}^{N}X_i - 2\sum_{i=1}^{N}Y_iX_i = 0$$

Then solution to the normal equations gives the least square estimates of *a* and *b* as follows:

$a = M_Y - bM_X$ where $M_Y$ and $M_X$ are the means of Y and X, respectively and:

$$b = \frac{\sum_{i=1}^{n}(Y_i - M_Y)(X_i - M_X)}{\sum_{i=1}^{n}(X_i - M_X)^2}$$

While the formula for the best-fitting line, $y = bx + a$, is same as one used to find a line in algebra, the points in this case don't lie perfectly on the line. The line is rather a model around which the data lie if a strong linear pattern exists.

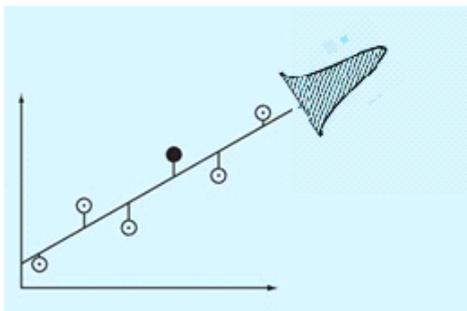

Fig. 3 Standard Error of Estimate

It is always helpful to assess the goodness of the fit. This is provided by the standard deviation of regression line, known as Standard Error of Estimate, Fig 3. It is given as: $S_{y/x} = S_r/N - 2$ where $S_r$ is the sum of squares of residuals, with N the number of data, and (N-2) the degrees of freedom.

It is not always that a linear relationship exists between the variables. In that case a linearization procedure is necessary. A sample illustration is given below:

$y = \alpha_1 e^{\beta_1 x} \rightarrow \ln y = \ln \alpha 1 + \beta 1 x$

$y = \alpha_2 x^{\beta_2} \rightarrow \log y = \log \alpha 2 + \beta 2 \log x$

## II. GENERAL LEAST SQUARES

It may be observed from preceding discussions that there are linear, polynomial and sinusoidal or other forms of equations with each apparently requiring a tailored set of equations. Actually, there is a general representative equation, as given below:

$y = b_0x_0 + b_1x_1 + b_2x_2 + \cdots + b_mx_m + e$   --- (1)

Note that $x_0, x_1, x_2, \cdots x_m$ are $m + 1$ basis functions. So where $x_0 = 1$, and $x_m = 0$ for $m > 1$, a simple linear regression equation obtains, otherwise it becomes a case of multiple linear equation. Furthermore, the expression defaults to polynomial form if the basis functions are simple monomials, such as when $x_m \Rightarrow x^m$ [3].

### A. Observation Equations

Observations equations are a form based on the general model outlined in (1), but as an adaptation to suite fields of application. In this instance, the interest is in horizontal networks in Geomatics Engineering. Therefore, whereas in laboratory experiments involving scatter plot diagrams, the quest is for an explanatory model, the case in surveying can be stated as follows:

1. The functions relating observables, angles and distances, to unknown positions, Easting and Northing, are known through geometry.
2. The task is to determine the unknown positions by the most optimal and reliable method.
3. The best solution is that for which the sum of squares of the residuals is a minimum.

In this context, an equation is formed for each measurement. This comprises of distance, height and angle observations, or enough to determine unknown variables of Easting (X), Northing (Y), and Height (Z). The resulting system of equations are usually such that, for *m* unknowns there are preferably *n* observations where n > m. As illustrated earlier in Fig. 3, this leads to a more reliable estimate.

Survey measurements are stochastic variables, with error considerations. As such, error propagation is a common issue, as regards its effect on the computed positions. Thus, let:

$$Y_1 = f_1(X_1, X_2, \ldots X_m)$$
$$Y_2 = f_2(X_1, X_2, \ldots X_m)$$
$$\vdots$$
$$Y_n = f_n(X_1, X_2, \ldots X_m)$$

where $Y_n$ is a function of variables, $X_1, X_2, \ldots X_m$.

The Jacobian matrix for these equations is defined as:

$$\mathbf{J_{yx}} = \begin{pmatrix} \frac{\partial Y_1}{\partial X_1} & \frac{\partial Y_1}{\partial X_2} & \cdots & \frac{\partial Y_1}{\partial X_m} \\ \frac{\partial Y_2}{\partial X_1} & \frac{\partial Y_2}{\partial X_2} & \cdots & \frac{\partial Y_2}{\partial X_2} \\ \vdots & \vdots & & \vdots \\ \frac{\partial Y_n}{\partial X_1} & \frac{\partial Y_n}{\partial X_2} & \cdots & \frac{\partial Y_n}{\partial X_m} \end{pmatrix}$$

If the covariance matrix of **x** is

$$\mathbf{C_x} = \begin{pmatrix} \sigma_{x_1}^2 & \sigma_{x_1 x_2} & \cdots & \sigma_{x_1 x_m} \\ \sigma_{x_2 x_1} & \sigma_{x_2}^2 & \cdots & \sigma_{x_2 x_m} \\ \vdots & \vdots & & \vdots \\ \sigma_{x_m x_1} & \sigma_{x_m x_2} & \cdots & \sigma_{x_m}^2 \end{pmatrix}$$

then the covariance matrix of y is: $C_y = J_{yx} C_x J_{yx}^T$ where

$$\mathbf{C_y} = \begin{pmatrix} \sigma_{y_1}^2 & \sigma_{y_1 y_2} & \cdots & \sigma_{y_1 y_n} \\ \sigma_{y_2 y_1} & \sigma_{y_2}^2 & \cdots & \sigma_{y_2 y_n} \\ \vdots & \vdots & & \vdots \\ \sigma_{y_n y_1} & \sigma_{y_n y_2} & \cdots & \sigma_{y_n}^2 \end{pmatrix}$$

Computations in Surveying Engineering involve linear models, **y = Ax**. In this case, the Jacobian matrix $\mathbf{J_{yx}}$ defaults to coefficient matrix, **A**. Hence

$$\mathbf{C_y = A C_x A^T} \qquad \text{---- (2)}$$

It is not always the case that information regarding the variance of observed quantities is available or superior to other forms of judgement. For instance ten rounds of observation would be more reliable than a single measurement. So weights, $w_i$ offer another option. The relationship may be expressed as follows:

$\sigma^2 \, \alpha \, \frac{1}{w}$, $\sigma^2 = \frac{k}{w}$. If $w = 1$ then $k = \sigma_0^2$, the variance of a measurement of unit weight.

$\therefore w = \frac{\sigma_0^2}{\sigma^2}$ and $\mathbf{W} = \sigma_0^2 . \mathbf{C^{-1}}$ ---- (3)

where **W** is the weight matrix.

Define Q, the cofactor as $\mathbf{Q} = \sigma_0^{-2} . \mathbf{C}$ ---- (4)

then pre-multiplying both sides of (3) by C gives

$$\mathbf{CW} = \sigma_0^2 . \mathbf{I}$$

And substituting for C in (4) gives the result:
**QW = I**.

Hence $\qquad \mathbf{Q = W^{-1}}$ ---- (5)

### B. Model Solutions

A model solution draws from the discussions so far, where the weight matrix **W** is a factor in determining the variance of computed positions.

Consider a situation where a linear mathematical model is represented by:

$$\mathbf{\bar{Y} = A\bar{X} + C} \qquad \text{---- (6)}$$

with $\bar{Y}$ the adjusted observations, A the coefficient matrix, $\bar{X}$ the unknowns and C, constants in the equation. In reality, $\bar{Y} = Y + \vartheta$ where Y are measurements and $\vartheta$ are residual corrections. Eqn (6) then assumes the form of:

$$Y + \vartheta = \mathbf{A\bar{X} + C} \qquad \text{---- (7)}$$

It is always helpful to reduce the magnitude of unknown quantities to improve computing performance. This involves the introduction of provisional and approximate values, such that

$\bar{X} = X_0 + X$. Substituting into (7) gives

$Y + \vartheta = \mathbf{A}(X_0 + X) + \mathbf{C}$. Hence collecting all,

$\vartheta = \mathbf{A}X - (Y - \mathbf{A}X_0 - \mathbf{C})$. Therefore the reduced equation with unknown corrections X is:

$$\mathbf{\vartheta = Ax - b} \qquad \text{---- (8)}$$

The best estimate of solution is obtained when the sum of squares of the residuals, $\vartheta = \mathbf{v^T W v}$, is at a minimum. Substituting for $\vartheta$ in (8) and proceeding thus [4]:

$$\begin{aligned}\vartheta &= \mathbf{(Ax - b)^T W(Ax - b)} \\ &= \mathbf{(x^T A^T - b^T) W(Ax - b)} \\ &= \mathbf{x^T A^T W Ax - x^T A^T W b - b^T W Ax + b^T W b} \end{aligned}$$

Therefore

$\partial \vartheta / \partial x = A^T W A x + (A^T W A)^T x - (b^T W A)^T - A^T W b$

$= 2 x^T A^T W A - 2 b^T W A = 0$

Hence, transposing: $A^T W A x = A^T W b$ ---- (9)
and
$x = (A^T W A)^{-1} A^T W b$ ----(10)

This result is the solution of the linearised model of the least squares equation. By equation (2) and [5], it can be shown that the variance of x is given by:

$\sigma_x^2 = \sigma_0^2 \, (A^T W A)^{-1}$ ----(11)

where $\sigma_0^2 = v^T W v/(n-m)$, n being the number of observations and, m the range of unknowns, n > m.

### III. FIELD APPLICATIONS

Field surveys involve establishing a control framework for engineering design and setting out of points to plan. As such high precision is required to ensure accurate design and compliance during construction. This can be achieved by adopting rigorous survey techniques and computational methods.

#### A. Linear Process

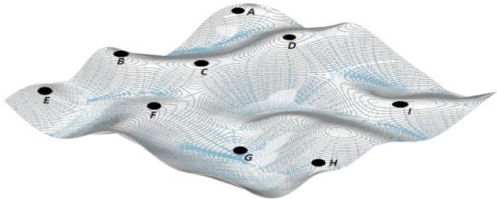

Fig. 5 Topographic Stations

A typical survey is illustrated above, where the stations are the centre for collecting grid point information in representation of an undulating topographic surface. The stations are computed by least squares, therefore requiring a linear model for implementation. This can be done by applying Taylor's theorem, expanding the function to first order approximation [6]:

$f(x) = f(x_0) + J \Delta x$ ---(12)

where $x_0$ is approximate value, $\Delta x$ the corrections and J the Jacobian matrix.

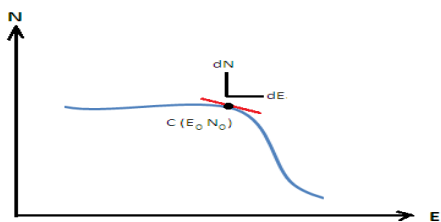

Fig.6 Linear Patch at C

Considering the above cross-section for point C, a linear patch is centred on the provisional coordinate $E_0\ N_0$. Then corrections dE and dN are solved by least squares to obtain $E_C = E_0 + dE$, and $N_C = N_0 + dN$. It is for this reason that provisional positions should be good enough to reduce the number of iterations.

1. *Basic Equations:*

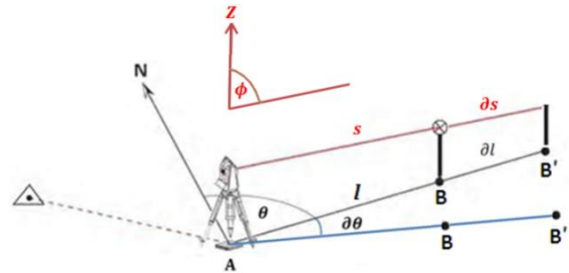

Fig.7 Position Fixing

Basic Equations can be demonstrated with Fig. 7, illustrating the process of fixing a station by angle and distance observations. The mathematical relationships are as follows:

$\theta = \text{Tan}^{-1}(\Delta E / \Delta N)$ ---- (13)

$l^2 = \Delta E^2 + \Delta N^2$ ---- (14)

Recalling Taylor's expansion in (12), then for $\theta = f(E_B, N_B, E_A, N_A)$, a linear model is:

$\theta = \theta_0 + (E_B - E_{B0})\dfrac{\partial \theta_{AB}}{\partial E_B} + (N_B - N_{B0})\dfrac{\partial \theta_{AB}}{\partial N_B} +$

$(E_A - E_{A0})\dfrac{\partial \theta_{AB}}{\partial E_A} + (N_A - N_{A0})\dfrac{\partial \theta_{AB}}{\partial N_A}$

Let $d\theta = \theta - \theta_0$ and $dE = E - E_0$ then after simplifying [7]:

$d\theta = \dfrac{Cos\theta}{l}(dE_B - dE_A) - \dfrac{Sin\theta}{l}(dN_B - dN_A)$ -(15)

Similarly, for $l = f(E_B, N_B, E_A, N_A)$

$dl = Sin\theta(dE_B - dE_A) + Cos\theta(dN_B - dN_A)$ -(16)

In respect of usage, the parameters are defined as follows:

$d\theta, dl$    Differences between the observed and computed values, with the latter derived from provisional position

$dE, dN$    Unknown corrections to provisional values of coordinates

The equations can be further simplified as:

*International Journal of Mathematics Trends and Technology (IJMTT) – Volume 37 Number 3 September 2016*$$dl = r(dE_B - dE_A) + s(dN_B - dN_A) \quad \text{-- (17)}$$
$$d\theta = p(dE_B - dE_A) - q(dN_B - dN_A) \quad \text{-- (18)}$$

where $p = \frac{Cos\,\theta}{l}$ and $q = \frac{Sin\,\theta}{l}$. Similarly, $r = Sin\theta$ and $s = Cos\theta$. Then taking into account the stochastic nature of the variables, the residual component $v$ is added to obtain the standard observation equations:

$$v_l = r(dE_B - dE_A) + s(dN_B - dN_A) - dl \quad \text{-- (19)}$$
$$v_\theta = p(dE_B - dE_A) - q(dN_B - dN_A) - d\theta \quad \text{-- (20)}$$

A derivative for angle observation, $\angle BAC$ equation from (20) is:

$$v_{BAC} = \{p(dE_C - dE_A) - q(dN_C - dN_A) - d\theta\}_{AC} - \{p(dE_B - dE_A) - q(dN_B - dN_A) - d\theta\}_{AB}$$
$$\text{--- (21)}$$

Hence
$$v_{BAC} = (p_{AB} - p_{AC})dE_A + (q_{AC} - q_{AB})dN_A - p_{AB}dE_B + q_{AB}dN_B + p_{AC}dE_C - q_{AC}dN_C - d\theta_{BAC}$$
$$\text{--- (22)}$$

Equations (19) and (22) form the basis of computation for horizontal positions in precision surveys. There is the option of 3D computation, but most often levelling provides superior height information in a separate network. However, it is a good choice to have default configurations, by providing for variables *s* and **φ** (Fig. 7) in support of 3D computations.

### *B. Examples*

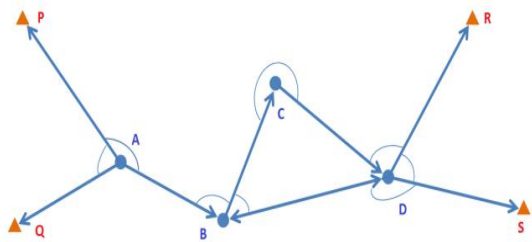

Fig. 8 Traverse Survey

A typical traverse survey involving angle and distance measurements is illustrated in Fig. 8, where P, Q, R and S are control stations and A, B, C and D are unknown stations. As such there are eight unknown parameters ($E_i$, $N_i$) and sixteen observations. A best fit solution is desirable.

1. *Coefficient Matrix:*

A coefficient matrix is the starting point in computation, representing the formation of least squares equations. Proceeding according to Eqn. (19) and (22), the matrix is constructed as follows:

| OBS | $dE_A$ | $dN_A$ | $dE_B$ | $dN_B$ | $dE_C$ | $dN_C$ | $dE_D$ | $dN_D$ |
|---|---|---|---|---|---|---|---|---|
| QAP | $p_{AQ}-p_{AP}$ | $q_{AP}-q_{AQ}$ | | | | | | |
| PAB | $p_{AP}-p_{AB}$ | $q_{AB}-q_{AP}$ | $p_{AB}$ | $-q_{AB}$ | | | | |
| ABC | $-p_{BA}$ | $q_{BA}$ | $p_{BA}-p_{BC}$ | $q_{BC}-q_{BA}$ | $p_{BC}$ | $-q_{BC}$ | | |
| CBD | | | $p_{BC}-p_{BD}$ | $q_{BD}-q_{BC}$ | $-p_{BC}$ | $q_{BC}$ | $p_{BD}$ | $-q_{BD}$ |
| BCD | | | $-p_{CB}$ | $q_{CB}$ | $p_{CB}-p_{CD}$ | $q_{CD}-q_{CB}$ | $p_{CD}$ | $-q_{CD}$ |
| CDR | | | | | $-p_{DC}$ | $-q_{DC}$ | $p_{DC}-p_{DR}$ | $q_{DR}-q_{DC}$ |
| RDS | | | | | | | $p_{DR}-p_{DS}$ | $q_{DS}-q_{DR}$ |
| SDB | | | $p_{DB}$ | $-q_{DB}$ | | | $p_{DS}-p_{DB}$ | $q_{DB}-q_{DS}$ |
| AQ | $-r_{AQ}$ | $-s_{AQ}$ | | | | | | |
| AP | $-r_{AP}$ | $-s_{AP}$ | | | | | | |
| AB | $-r_{AB}$ | $-s_{AB}$ | $r_{AB}$ | $s_{AB}$ | | | | |
| BC | | | $-r_{BC}$ | $-s_{BC}$ | $r_{BC}$ | $s_{BC}$ | | |
| BD | | | $-r_{BD}$ | $-s_{BD}$ | | | $r_{BD}$ | $s_{BD}$ |
| CD | | | | | $-r_{CD}$ | $-s_{CD}$ | $r_{CD}$ | $s_{CD}$ |
| DR | | | | | | | $-r_{DR}$ | $-s_{DR}$ |
| DS | | | | | | | $-r_{DS}$ | $-s_{DS}$ |

Table 1

2. *Computation:*

Computation based on the resultant equation is: $v_{16,1} = A_{16,8}X_{8,1} - b_{16,1}$. It involves a number of stages:

i. Collation
ii. Classification
iii. Data Input
iv. Computation

Collation involves compiling observations and measurements into a single list, while in classification, a decision is made as regards fixed points in the network.

In data input, the construction of the coefficient matrix and equations is implicit, following the guidelines of network diagram. In non-static situations, this mode is prone to misrepresentation and errors. And so, the results of computation would always depend on the correctness in the three stages, as afore-mentioned.

### IV. AUTOMATION

The stages listed above are obviously manual and was the case at the early stages of computer applications. Yet, even at present, this modality still persists in many instances, ported into computer solutions. What is most desirable is a fully automated approach, robust and able to cope with dynamic networks.

ISSN: 2231-5373                    http://www.ijmttjournal.org                    Page 194

### A. Frames

The concept of frames as discussed widely in [8] is one of knowledge representation. With origins in Artificial Intelligence this notion has evolved into what is generally regarded as object oriented programming.

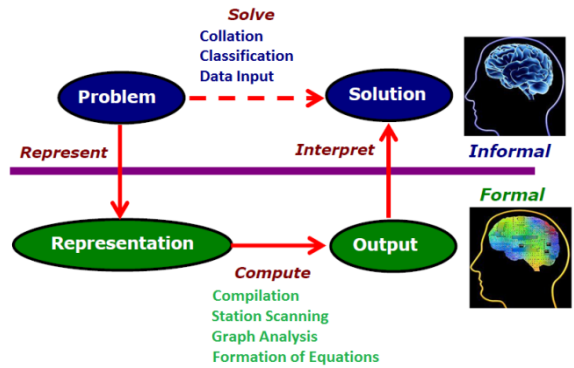

Fig. 9 KR in Least Squares Computation

The process outlined earlier is an informal approach as would a human expert. But in Fig. 9 is an illustration of the relationship as how this expertise could be represented and translated into a formal expression that is understood by the machine.

It follows therefore that the stages in manual approach can be replaced by a single instruction in frame-based technique and object oriented programming [9].

#### 1. Compilation:

Compilation features three main interacting nodes, namely Compile, Station Description and Traverse Field Book. The connexion is best described in a semantic net [10], [11], as in Fig. 10.

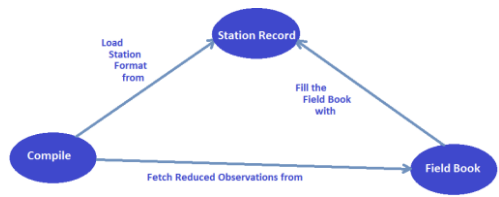

Fig. 10 Semantic Net

A starting node is the station record that provides format and details of information regarding observations from instrument station. The field book, on the other hand is a collection or container database with contents of all the observations and measurement in the survey.

Compile fetches the field book and along with the station data structure, generates mean values of reduced measurements to create a new data set. The latter becomes the input source in subsequent computations.

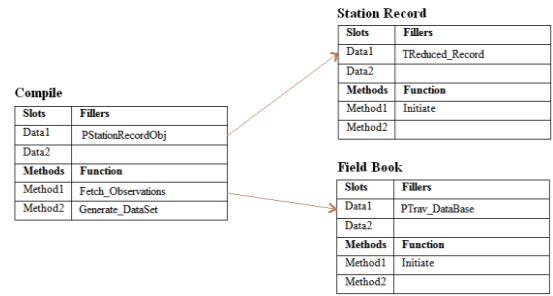

Fig. 11 Compile Frame

The semantic net at a glance is a useful summary, but does not provide full details. To go further requires frame representation as in Fig. 11. The slots and fillers are apparent as data fields and types. Then the advantage of methods is obvious, as being able to interact with other frames and execute routines. Thus, in the second method, Generate DataSet carries out the exercise after loading the field book.

#### 2. Station Scanning:

Station scanning is a process that amongst others identifies fixed points in the network. There are therefore three contending frames, namely Control Database, Complied Dataset, and Fixed Stations in the Dataset.

The concept of Control database draws from manual administration where there is always a register of known Trig/Control points. Usually before the project begins, decisions are made as to which of these points will serve in the project. At times, also fixed stations are chosen on the fly and later admitted for computation. In all cases controls are defined with respect to a particular datum and there are always sufficient overlaps to allow result computed in one datum to be listed in another.

In the fore going therefore, all known fixed points are listed in the control database. It is a simple matter of frame intersection search to identify them in the dataset, Fig. 12.

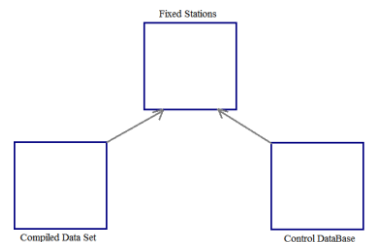

Fig. 12 Intersection Search

Fixed Station Set = DataSet∩Control DataBase

Once this is accomplished, the class of the stations are updated as fixed points, otherwise a free status becomes applicable.

### 3. *Graph Analysis:*

Graph Analysis provides a description of the network, an analogy to field diagram. Often designed networks are not followed in the field and it is in this context that this analysis, especially in large dynamic networks, provides a reliable on-the-field report.

With respect to traverse survey, Fig. 8, the analysis is based on theory of directed graph, where thenodes and edges complete the description.

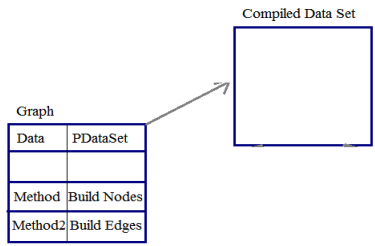

Fig. 13 Graph Frame

A typical frame is illustrated in Fig. 13, where interaction with Dataset generates reference list of observations in PDataSet. Methods 1 and 2, then provide the description.

In observation equations model, Graph Analysis is not critical for computations itself, but it does provide the basis for processing cycles and closures, conditions in the network.

### 4. *Formation of Equations:*

Formation of equations is the culminating process setting the stage for computation. It is a routine embedded within the Least Squares frame.

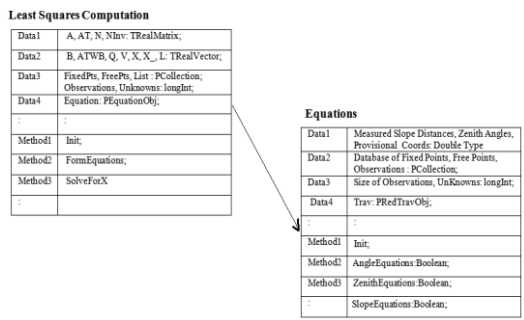

Fig. 15 Computation Frame

The computation begins with a call through Method2 to Equations frame, by way of Data4, for each observation. Itis processed to fill the row for that observation, depending on either of Angle, Distance or Zenith measurement. Upon return, the data is used to update the A, B and W matrices defined in Data 1 and 2.And following the equations, Method3 provides solution in $\mathbf{x} = (\mathbf{A}^T\mathbf{WA})^{-1}\mathbf{A}^T\mathbf{Wb}$.

### B. *Cycle Processing*

The essence of cycle processing is to generate error information to determine internal and external consistency in the network. By description a graph maybe defined as acyclic, meaning there are no cycles [12]. However in context of survey applications, the emphasis is on adaptation, recognition of geometry and topology. Therefore, a cycle or loop starts from a node, passing through other nodes and terminating at the start node. Thus, where a DAG exists, the entities can be reduced to cycles by reversing observed directions in affected segments.

Processing cycles starts with a spanning tree and addition of cross or back edges. The key algorithms are Breadth First Search and Depth First Search [13].

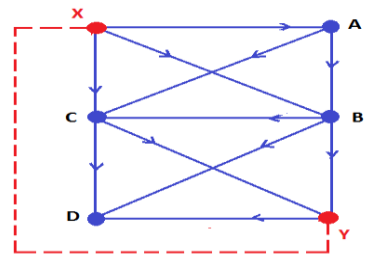

Fig. 14

In Fig 14 is a model diagram of a field survey. Typically, work originates from a control point and terminates like wise. Thus, observed directions are important in computations and are reflected in processing algorithms. With respect to the model, the survey involves transfer of position from X, proceeding and closing at Y. Further observations allow determination of C and D, to complete the network as a connected graph.

#### 1. *Breadth First Search:*

Breadth First Search starts by investigating each direction emanating from X, the first station. It then moves to next layer, the next station B until finally all the nodes are connected in a spanning tree, Fig 16.

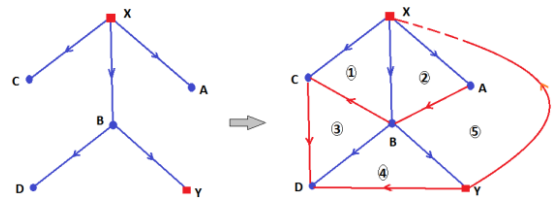

Fig. 16

Next fundamental cycles are constructed by adding cross edges. There are 5 cycles where the fifth, XBYX ensures consistency with external framework.

### 2. *Depth First Search:*

Depth First Search, operating from the first station X follows a one-way route in the main direction of survey, until the last stop D, when there is no further path to proceed. It then retraces its way backwards to investigate unexplored routes, BC, until completion that leads to a spanning tree, Fig 17.

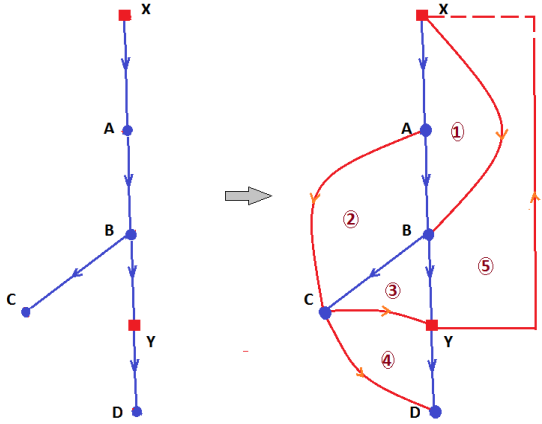

Fig. 17

Back edges are then fitted to construct fundamental cycles, in order to establish the condition equations. These cycles are: XABX, ABCA, BYCB, YDCY, and XBYX.

### 3. *Model Solution:*

A model solution arises out of comparison of the two search options, BFS and DFS. The consensus of opinion is that choice depends on the nature of data structure. In this regard, a survey route is characterized by depth and DFS is a natural representation. Furthermore, it does not entail huge memory requirements.

BFS nevertheless has advantage of minimum path length in cycle formation. Involving cross edges, the cycles have less overlap and agrees more readily with visual observations of geometry. However it does not lend with run of survey and topology.

Generally, as described in [14], the best option is offered by Depth - Limited Search, which is to say that the finest solution lies in improving DFS. Thus, in this application, DFS is adopted and have proved very robust and reliable.

### V. PROGRAMMING

In programming, frames give way to models in object oriented design and an example of which may be found in Pascal language such as, Lazarus [15] and Delphi [16]. The Least Squares frame now assumes the form description below.

```
PLsqAdj = ^TLsqAdj;
TLsqAdj = object(TObject)
A,AT,N,NInv:TRealMatrix;
 B,ATWB,Q,V,X,X_,L:TRealVector;
 Equation: PEquationObj;
SightDist,UnitVariance,SightVar:double;
   :             :              :
FixedPts,FreePts,List : PCollection;Dim:integer;
FixedSet,FreeSet,DataSet: array[0..70] Of Char;
Obs, UnKnws: longInt;;
constructorInit;
   :             :              :
procedureFormEquations;
procedureSolveForX;
destructor Done; virtual;
end;
```

It may be noted, the complement between the two models. While frames offer design basis, object models provide the required platform for implementation.

### A. Algorithm

The algorithms address formation of equations and generation of errors for each cycle in the network.

### 1. *Formation of Equations:*

The procedure FormEquations is critical in its role of constructing the coefficient matrix and equations. The program steps for modelling this formation can be described with respect to the Survey in Fig 8.

| STN | A | | B | | C | | D | |
|---|---|---|---|---|---|---|---|---|
| INDEX | 1 | | 2 | | 3 | | 4 | |
| ΔX | $dE_A$ | $dN_A$ | $dE_B$ | $dN_B$ | $dE_C$ | $dN_C$ | $dE_D$ | $dN_D$ |
| COL | 1 | 2 | 3 | 4 | 5 | 6 | 7 | 8 |

Fig. 18

1. Form a list of free stations in the network and store in a vector, $STN_{1xN}$, Fig 18.
2. Assign index, sequence of numbers, to each station in the vector.
3. Create another vector, the size of unknowns.
4. Then for any observation involving a station, say C in BĈD, locate the station and index.
5. Multiply the index by dimension, 2 or 3, to compute column positions in the row of coefficient matrix: ∴ Col = 2 x Index
6. Store the coefficient for $dE_C$ at: Col -1.
7. Store the coefficient for $dN_C$ at: Col.
8. Do the same for B and D, and Return to 4.

In the foregoing, with respect to Fig. 8 and Table 1, it can be seen that there is agreement in columns 5 and 6 for coefficients of $dE_C$ and $dN_C$.

### *2. Generation of Cycles:*

In earlier discussions, the DFS was demonstrated as a more suitable algorithm for cycle processing. However, for implementation the mechanism differs from standard approach involving stacks etc. Rather what is adopted in this instance is an alternative to suit survey applications.

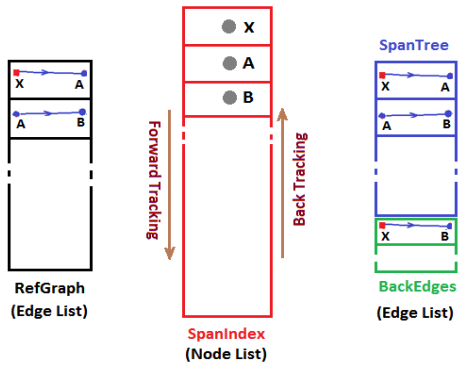

Fig. 19 Program Steps

With respect to Fig. 6 representing the network, a program implementation is described with illustrations in Fig 19. Thus, RefGraph is a variable representation for the structure as edge list. The same form applies to SpanTree and BackEdges. SpanIndex is a node list.

The process starts with initialization where SpanIndex, SpanTree are empty. The first segment XA, invariably the origin of the survey is evaluated and inserted into SpanTree. At the same time, the nodes X and A are inserted into SpanIndex. The target station A becomes the reference in forward tracking that yield the edge AB. Both the SpanIndex and SpanTree are subsequently updated. Forward tracking continues until a leaf, D, is encountered. Back tracking then follows until B is accessed and a new segment BC is found and explored, Fig 17.

Completion produces a DFS spanning tree. The back edges are generated, and form the basis for constructing fundamental cycles and condition equations in the network. Thus, by so doing, provides the determination of internal and external errors, consistencies in the survey.

### *B. Optimisation*

The purpose of optimisation is to harness all the frames or instructions into a minimum, to achieve high productivity. To this end, there are a number of options to consider.

### *1. Frame Integration:*

In the discussions so far, the impression might be that each of the frames have to be run separately. Well, except where it is necessary to carry users along in the progress of computations, by pause or interface enquiries, this should not be the case.

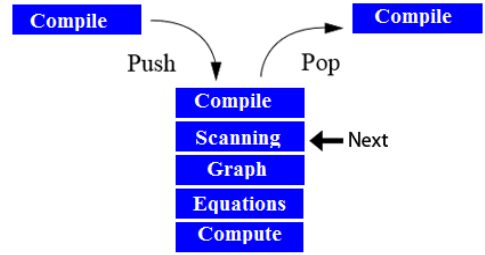

Fig.20 Frames in Stack

Ideally, the software should interact with field data, reach conclusions and then compute results in one instruction. This can be done by the starting frame, Compute, invoking required routines as illustrated in a PUSH and POP arrangement, Fig. 16.

### *2. Datum Transformation:*

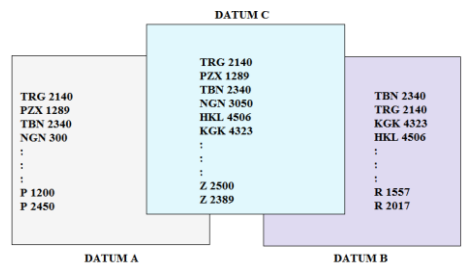

Fig.21 Datum in Frames

In the diagram above are control points in their respective datum. It is common to carry out a project in A, for instance, and supply the results in C. This is mostly an office activity, involving Least Squares estimation.

Fig. 22 Results List

It would be best though to compute as in 1, and simply list results in any desired datum as in Fig. 22.

## VI. CONCLUSION

This paper has reviewed the principle of Least Squares estimation from its origin in Statistics, Regression Analysis. And further, onto the adaptation of its use in Geomatics Engineering. It is evident that Least Squares computation is the optimal method for getting best results. However, adopting this technique has always been fraught with challenges.

The approach discussed in this paper has resolved these issues with automation. So save input activity and registration of control points, the process is straight forward. Furthermore, with automated facilities for error analysis, to compute work in one datum and list results in another, entire computation including reduction of detail points, can now be carried out in the field rather than office.

On the whole, it is another step towards the goal of field-to-finish in productivity.